\documentclass[useAMS,usenatbib]{mn2e}
\usepackage{textcomp}
\usepackage[utf8]{inputenc}
\usepackage{threeparttable}
\usepackage{epsfig}
\usepackage{graphicx}
\usepackage{amssymb,amsmath}
\usepackage{hyperref}
\usepackage{mn2e-bst}
\newcommand{\apj}{ApJ}
\newcommand{\aap}{A\&A}
\newcommand{\araa}{ARA\&A}
\newcommand{\apjl}{ApJL}
\newcommand{\mnras}{MNRAS}
\newcommand{\apss}{Ap\&SS}
\newcommand{\pasj}{PASJ}
\newcommand{\pasp}{PASP}

\title{High-energy Emission Processes in M~87}

\author[de Jong et al.]
{S.~de Jong$^{1}$,   
V.~Beckmann$^{1}$, 
S.~Soldi$^{1}$, 
A.~Tramacere$^{2}$ 
\& A.~Gros$^{3}$\\
$^{1}$Fran\c{c}ois Arago Centre, APC, Universit\'e Paris Diderot, CNRS/IN2P3, CEA/Irfu, Observatoire de Paris, \\
Sorbonne Paris Cit\'e, 10 rue Alice Domon et L\'eonie Duquet, 75205 Paris Cedex 13, France \\
$^{2}$ ISDC Data Centre for Astrophysics, 16 ch. d'\'Ecogia, 1290 Versoix, Switzerland\\
$^{3}$ Service d'Astrophysique / IRFU / CEA - Saclay, 91191 Gif sur Yvette, France}

\begin{document}

\maketitle

\begin{abstract}
 
We study the contribution of thermal and non-thermal processes to the inverse Compton emission of the radio galaxy M~87 by modelling its broad-band emission. Through this we aim to derive insight into where within the AGN the X-ray, $\gamma$-ray, and VHE emission is produced.\\
We have analysed all available {\it INTEGRAL} IBIS/ISGRI data on M~87, spanning almost 10 years, to set an upper limit to the average hard X-ray flux of $f(20 - 60 \rm \, keV)\la 3\times 10^{-12}$  $\rm \, erg \, cm^{-2} \, s^{-1}$, using several techniques beyond the standard analysis which are also presented here. We also analysed hard X-ray data from {\it Suzaku}/PIN taken late November 2006, and we report the first hard X-ray detection of M~87 with a flux of $f(20 - 60 \rm \, keV) = 10^{-11}\rm \, erg \, cm^{-2} \, s^{-1} $. In addition we analyse data from {\it Fermi}/LAT, {\it INTEGRAL}/JEM-X, and {\it Suzaku}/XIS. We collected historical radio/IR/optical and VHE data and combined them with the X-ray and $\gamma$-ray data, to create broad-band spectral energy distributions for the average low-flux state and the flaring state. The resulting spectral energy distributions are modelled by applying a single-zone SSC model with a jet angle of $\theta=15\,^{\circ}$.  We also show that modelling the core emission of M~87 using a single-zone synchrotron self-Compton model does represent the SED, suggesting that the core emission is dominated by a BL Lac type AGN core. Using SED modelling we also show that the hard X-ray emission detected in 2006 is likely due to a flare of the jet knot HST-1, rather than being related to the core.\\

\end{abstract}


\section{Introduction}
Radio galaxies are a subclass of AGN, which display jets that are observed with a large angle of $\theta>10^{\circ}$ with respect to the line of sight, enabling a view of both the jet and the core. In the unification model of AGN, they are thought to be the radio-loud counterparts of Seyfert galaxies \citep{Antonucci1993,Urry1995,2012agn}. 
In the case of blazars $\gamma$-ray emission is expected to be detected due to the small jet angle and the resulting relativistic beaming towards the observer. The relativistic jet will dominate the emission, and due to Doppler boosting this emission can reach into the $\gamma$-ray and TeV range. However, several non-blazar AGN and in particular radio galaxies have also been detected in the $\gamma$-rays \citep[see for example the third {\it Fermi}/LAT source catalogue;][]{thirdsourcecat2015}, and a few of these sources have been observed at energies $>100 \rm \, GeV$ \citep{Perkins2012}. The mechanism driving this high-energy emission in radio galaxies is still under discussion.

M~87 is a FR-I radio galaxy \citep{FanaroffRiley1974MNRAS,Laing1983}, with a central supermassive black hole of mass \mbox{$\rm M_{BH}=(3-6)\times 10^{9}\, \rm M_{\odot}$} \citep{Macchetto1997,Gebhardt2009, Batcheldor2010}, at a distance of 16 Mpc \citep{Tonry1991}. The jet angle has been estimated at $\theta=15^{\circ}$ with respect to the line of sight based proper motions of the jet features \citep{Biretta1999}. 

M~87 has been detected by {\it Fermi}/LAT \citep{Abdo2009}, making it the third radio galaxy to be detected in $\gamma$-rays, after Centaurus~A and NGC~1275. M~87 has also been detected in the VHE range during flares, e.g. by {\it HEGRA} \citep{Aharonian2003}. This source shows variable emission in the soft ($< 10\rm \, keV$) X-ray regime \citep{Harris2009}, and extrapolating from high flux states observed by {\it Chandra} ($f \simeq 2 \times 10^{-12} \rm \, erg \, cm^{-2} \, s^{-1}$), a flux of \mbox{$f \simeq 10^{-12} \rm \, erg\, cm^{-2} \, s^{-1}$} would be expected between 20 and 60 keV. \citet{Walter2008} reported a detection with  
\mbox{$f=(8.6 \pm 1.8)\times 10^{-12} \rm \, erg \, cm^{-2} \, s^{-1}$} 
between 20--60 keV using {\it INTEGRAL}/ISGRI data. However, we will show that this detection can not be confirmed using the latest software and instrument calibration.

The proximity of M~87 allows us to image the core separately from the jet and the diffuse extended emission in several wavelengths like radio, optical, and soft X-rays, but not in hard X-rays ($> 10 \rm\, keV$) nor at higher energies. At energies above $10 \rm\, keV$ the resolution of commonly used coded-mask instruments (e.g. {\it INTEGRAL}, {\it Swift}/BAT) is of the order of 10'. 
Only recently with NuSTAR’s narrow field instrument a resolution of $\sim$ 10'' can be achieved at hard X-rays using grazing incidence optics.

By studying the spectral energy distribution (SED) it is possible to test whether the high-energy emission originates from the core, the jet, or from an extended region.
Earlier SEDs of M~87 have been represented by a synchrotron self-Compton (SSC) type model \citep[see for example][]{Abdo2009}. SSC models are often used to represent blazar spectra, but they have also been applied to other $\gamma$-ray detected radio galaxies, such as Centaurus~A \citep{AbdoCenA2010}. The latter study was inconclusive, though, on the question whether the X-ray domain is dominated by the non-thermal jet emission, or arises rather from a Seyfert type core \citep[see also][]{Beckmann2011}.

In this paper we present an upper limit on the average long-term hard X-ray emission of M~87 using 1.7 Ms of {\it INTEGRAL} IBIS/ISGRI data and different techniques for performance enhancement of {\it INTEGRAL} IBIS/ISGRI. We have also analysed {\it Suzaku} data from November 2006, where we have detected M~87 for the first time between 20--60 keV. This is combined with other data in soft X-rays provided by {\it INTEGRAL}/JEM-X, $\gamma$-rays by {\it Fermi}/LAT, historical radio, infrared and optical emission of the core, and VHE data from H.E.S.S. to create an average SED. To understand the 2006 {\it Suzaku} observations we create two simultaneous SEDs: one for the M~87 core and one for the bright jet knot HST-1.
Error values quoted in this paper are at the 1-$\sigma$ level unless indicated otherwise.

\section{Data analysis}

\subsection{\it INTEGRAL}
In this study we used all available data on M~87 taken by the {\it INTEGRAL} mission \citep{INTEGRALmission} since its launch. Observations have been performed in dithering mode with pointed observations, so-called science windows, lasting between 2000 s and 4000 s. The data cover the time from 2003 to 2011, and most of the observations are taken after 2008. We analysed data from the Joint European Monitor in X-rays (JEM-X) and from the IBIS/ISGRI imager.
JEM-X is a coded-mask instrument that consists of two identical co-aligned telescopes and operates in the 3 to 35 keV band \citep{Lund2003}. The field of view is circular with a diameter of 4.8 degrees (fully coded) and an angular resolution of 3.25 arcminutes.  
We created images in the energy ranges of 3--10 keV and 10--25 keV for each pointing and combined these into a single mosaic image for both JEM-X detectors. In the 3--10 keV energy band we detect M~87 with a significance of $15 \sigma$ and a flux of $f=1.6 \times10^{-11} \rm \, erg \, cm^{-2} \, s^{-1}$. In the 10--25 keV energy band M~87 is not detectable, with a $3\sigma$ upper limit of $f \la 1.2\times10^{-11} \rm \, erg \, cm^{-2} \, s^{-1}$.
 
\begin{figure}
 \includegraphics[width=8cm, keepaspectratio=true]{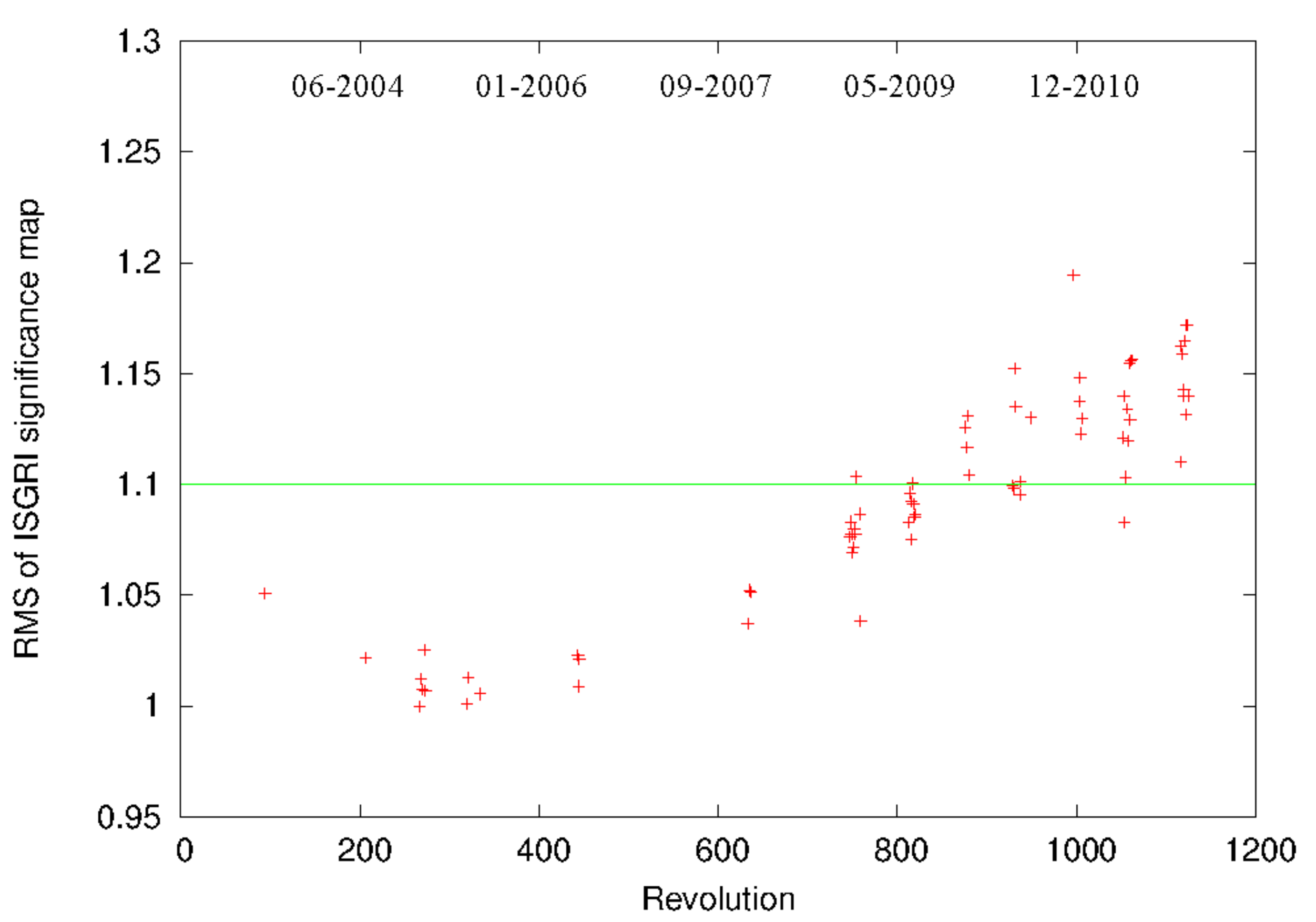}
 \caption{The evolution of the $s_{\rm rms}$ of the significance image of the mosaic per revolution. The line shows the cut for $s_{\rm rms}=1.1$. Later revolutions, starting around revolution 800 (2009), have on average a higher rms.}
 \label{mosaic_features}
\end{figure}

The {\it INTEGRAL} Soft Gamma-Ray Imager (ISGRI) is part of the Imager on Board {\it INTEGRAL} Spacecraft (IBIS), which is also a coded-mask instrument. ISGRI is sensitive between 15 keV and 1 MeV \citep{Lebrun2003}. The field of view is $9^\circ \times 9^\circ$ (fully coded), and the angular resolution, limited by the coded mask technology, is \mbox{12 arcminutes}. The total IBIS/ISGRI data set reaches an effective on-source exposure time of 1.7 Ms (see Fig.~\ref{mosaic_features}). Using the standard \mbox{Offline Scientific Analysis} (OSA) package version 9 provided by the ISDC \citep{Courvoisier03} we created images for each pointing in the energy range 20--60 keV and combined them into a mosaic. At the position of M~87 we derived a detection significance of $3.8 \sigma$ and a flux of \mbox{$f \simeq 3\times 10^{-12}$  $\rm \, erg \, cm^{-2} \, s^{-1}$}. However, the mosaic image shows strong noise features in the vicinity of M~87, and thus the detection cannot be deemed as trustworthy. To improve the quality of the image we applied several techniques, a summary of which can be found in Table~\ref{isgri_res}. 

In order to quantify the quality of the mosaic images, we determined  histograms of the significance image and the root mean square (rms, $s_{\rm rms}$) of this histogram, where in the ideal case the mean should be 0$\sigma$ and the $s_{\rm rms}=1$. Using the $s_{\rm rms}$ we can track the improvement of the mosaic quality. For the total mosaic created using OSA~9 we derived $s_{\rm rms}=1.75$ (see Tab.~\ref{isgri_res}).
To improve the image we started by excluding those science windows from the mosaic analysis that showed a high noise level, i.e. we removed all science windows with $s_{\rm rms} >1.2$ ($\sim 4\%$ of the total).  
The global image quality improved, and the M~87 detection is no longer significant. We have also processed the good science windows with the more recent OSA~10 software. We found a detection significance of $1.32 \sigma$ at the position of M~87 and derived an upper limit of $f \la 3.3 \times 10^{-12} \rm \, erg \, cm^{-2} \, s^{-1}$, within a mosaic with a rms of $s_{\rm rms}=1.38$ (OSA~10 in Table~\ref{isgri_res}). 

We then produced mosaics per revolution (a revolution lasts about 3 days and has an effective exposure time of $\sim 200 \rm \, ks$) to evaluate their quality based on the $s_{\rm rms}$ value of the significance map. For the mosaics we set a lower rms threshold of 1.1, because the fluctuations in the mosaics are more averaged out compared to the single science windows. Most of the revolutions that have $s_{\rm rms}<1.1$ are within the first 6.5 years of the mission (rev. $ < 800$, see Fig.~\ref{mosaic_features}). This is probably due to the evolution of the background (private communication).
We then combined the images from revolutions where the rms is $s_{\rm rms}<1.1$. The rms of the combined mosaic is $s_{\rm rms}=1.18$ and we derived for M~87 a $3\sigma$ upper limit of $f < 4.2 \times 10^{-12} \rm \, erg \, cm^{-2} \, s^{-1}$ (Tab.~\ref{isgri_res}: OSA 10 selected revolutions). We also made a mosaic with the combined revolutions with rms $s_{\rm rms}>$1.1, which resulted in an upper limit to the flux of $4.9 \times 10^{-12} \rm \, erg \, cm^{-2} \, s^{-1}$ and the mosaic rms of $s_{\rm rms}=1.39$.

Lastly we used a technique which required changes to the program {\tt ii\_shadow\_ubc}, which is part of the OSA 10 pipeline. This routine performs the background correction using the detector images (shadowgrams). To decrease the noise in the images we removed pixels from the borders of the shadowgrams, since the borders of the detector contain the most unpredictable instrumental background. 
After testing several configurations on single revolutions we used a cut of the 3 outmost pixels to create 3 mosaics containing all available observations: one mosaic contains the non-modified science windows (control), one mosaic where all science windows have been modified, and one mosaic where cuts were applied only for those revolutions with a high rms value. To quantitatively evaluate these mosaics we calculated the rms $s_{\rm rms}$ of the significance map, where we now considered only the inner $10^\circ \times 10^\circ$ 
since the borders of the mosaics are unnaturally smooth due to the cutting procedure of the shadowgrams, and the area around bright sources in the field is excluded. The rms for the mosaic with all science windows modified is the lowest, $s_{\rm rms}=1.40$, and for the unmodified mosaic the highest, $s_{\rm rms}=1.54$. 
However, we found that these techniques did not further alter the upper limit on the flux of M~87. 

In {\it INTEGRAL}/ISGRI observations the systematic errors are dependent on exposure time and on the number of sources in the field and their brightness. In the M~87 field the number of sources is relatively low and the sources present are only moderately bright, therefore limiting the influence of systematic errors. In addition, the systematic errors are averaged due to the broad energy range and large observation time. In this field it has been verified that the systematic errors are negligible compared to the statistical errors. 

\begin{table*}
\centering
\caption{Results of IBIS/ISGRI analysis on the field of M~87 applying different data selection criteria. The detection found with the OSA 9 analysis is spurious. The count rates have been converted into fluxes assuming a Crab-like spectrum. }
{\footnotesize
\begin{tabular*}{\textwidth}{lccccc}  
\hline \hline
\noalign{\smallskip}

  Method & Mosaic rms $s_{\rm rms}$ & Detection significance  & count rate & M~87 $3 \sigma \,$ upper limit& Detection significance \\
       &            & M~87 [$\sigma$]             &   [s$^{-1}$]     &  [$10^{-12} \rm \, erg \, cm^{-2} \, s^{-1}$] & NGC~4388 [$\sigma$]\\
\hline
\noalign{\smallskip}
OSA 9			       & 1.75 & 3.77 & 2.34$\pm$0.02 & 3.0 & 148.0 \\
OSA 10                        & 1.38 & 1.35 & 2.20$\pm$0.02 & 3.3 & 113.5\\
OSA 10: selected revolutions  & 1.18 & 1.70 & 2.33$\pm$0.03 & 4.2 & 89.4\\
OSA 10: borders cut           & 1.40 & 1.16 & 2.20$\pm$0.02 & 3.2 & 111.3\\
\noalign{\smallskip}
\hline 

 \end{tabular*}
\label{isgri_res} 
}
\end{table*}

\subsection{ {\it Suzaku}}

 {\it Suzaku} has observed M~87 from November 29 to December 2, 2006 with an elapsed time of 187~ks in HXD nominal pointing mode. We analysed data from both the X-ray Imaging Spectrometer \citep[XIS,][]{Koyama2007} and the Hard X-ray Detector \citep[HXD,][]{Takahashi2007}. 
The XIS instrument operates between $\sim$0.2--12.0 keV and consists of three separate CCD detectors with a field of view of 18' x 18'  and a spatial resolution of 1.6-2.0 arcmin. The XIS data were reprocessed applying the standard event cuts and a spectrum was extracted from a circular region with a 60'' radius around the core.  Due to calibration issues the observations between 0.4 to $\sim$3 keV show strong residuals. The 3--7 keV band displays thermal signatures such as a tentative iron K$\alpha$ line at 6.7 keV (upper limit to the equivalent width of 200 eV), indicating that the emission originates from a hot diffuse gas. This is due to the extraction region used that includes both the core of M87 and the surrounding hot gas. 
Since we are interested in the non-thermal core emission, we model only the 7--10 keV band, where the thermal emission from the hot gas is not dominant. An absorbed power-law model yielded a fit of $\chi_\nu^2=1.03$ (for 150 d.o.f.), with a fixed Galactic hydrogen column density $N_{\rm H}=2 \times 10^{\rm 20} \rm \, cm^{\rm -2}$ and a power law index of $\Gamma=2.3\pm0.3$ (90\% error). The extrapolated 2--10 keV power law flux is $f=(2.5^{+0.2}_{-0.4})\times 10^{-11}  \rm \, erg \, cm^{-2} \, s^{-1}$. 

\begin{figure} 
 \includegraphics[width=8cm, keepaspectratio=true]{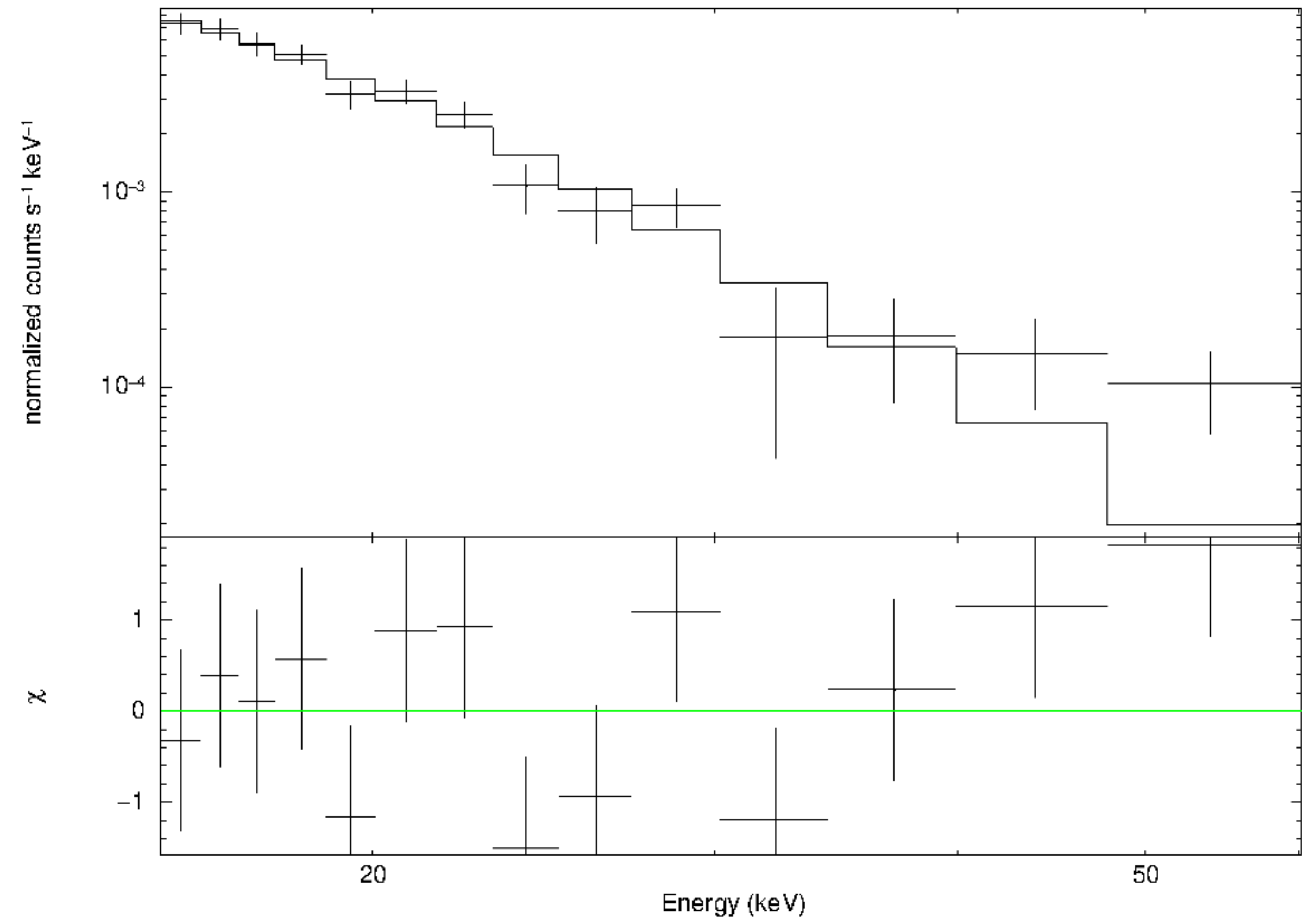}
 \caption{Suzaku/PIN count spectrum between 15 and 70 keV with an absorbed power law model fit. The bottom panel shows the residuals of the fit in terms of the standard deviation with error bars of size one sigma.}
 \label{pinspec}
\end{figure}

The HXD is a collimated detector with a field of view of 34' x 34'. The detector that consists of two independent systems: silicon PIN diodes that operate in the range $\sim$10--60 keV and GSO scintillation counters that function between $\sim$30--600$ \rm \, keV$. We reprocessed the data applying the standard event cuts. Due to the low count rate in the GSO band no significant detection could be extracted from this detector. For the PIN detector we extracted a significant spectrum between 15 and 70~keV (Fig.~\ref{pinspec}). The spectrum can be represented by an absorbed power law with a fixed column density $N_{\rm H}=2 \times 10^{\rm 20} \rm \, cm^{\rm -2}$ and a power law index of $\Gamma=2.8^{+0.5}_{-0.4}$ (90\% errors) giving a reduced $\chi_\nu^2=1.17$ (for 12 degrees of freedom). The model flux is $f = (1.04^{+0.03}_{-0.19})\times 10^{-11} \rm \, erg \, cm^{-2} \, s^{-1} $ between 20 and 60 keV. The combined XIS and HXD spectrum modelled with an absorbed power law shows a best fit of $\chi_\nu^2=1.03$ (150 d.o.f.) and a power law index of $\Gamma=2.6\pm0.2$ (90\%
 error). While the spectral indices of the XIS and HXD differ, they are consistent within the 90\% confidence intervals.  

\subsection{{\it Fermi}/LAT}

The Large Area Telescope \citep[LAT,][]{LAT}  aboard the {\it Fermi} satellite is a pair-conversion instrument that is sensitive between 20 MeV and 300 GeV. We have used all available data taken between August 2008 and May 2012, with a total effective exposure of 30 Ms. We have selected source-class-events (P7SOURCE\_V6) between 100 MeV and 100 GeV in a circular region of $30^{\circ}$ around M~87. The large extraction radius is necessary for the binned analysis we performed. Events with a zenith angle of more than $100^{\circ}$ were excluded, and we used the standard cuts proposed by the {\it Fermi} team based on the data quality of the events and on the instrument configuration. In order to get a reliable result for the source of interest, we included in the maximum likelihood fitting procedure all sources that had been reported in the second {\it Fermi} catalogue (2FGL) within $45 ^{\circ}$ around the position of M~87. 

Between 100 MeV and 100 GeV we derived a flux for M~87 of $f = (2.2 \pm 0.3)\times 10^{-8} \rm \, ph  \, cm^{-2} \, s^{-1}$ and a power law index of $2.16 \pm 0.06$, with a  test statistic $TS=370$, which corresponds to a detection significance of $\sim 19 \sigma$. Since the source is bright in $\gamma$-rays, we divided the energy range 100 MeV to 100 GeV into 5 logarithmic bins to track the spectral evolution with energy. The results are summarized in Table~\ref{fermi}, where we give the significance, flux and power law index per energy range.  

\begin{table}
\caption{Results of the binned likelihood analysis of the {\it Fermi}/LAT data in 5 bins between 0.1--100 GeV with a power law of photon index $\Gamma$. The errors are given at $1\sigma$ confidence level.}
{\footnotesize
\begin{tabular}{c c c r c}  
\hline \hline
\noalign{\smallskip}
  bin & energy range & flux & $TS$ & $\Gamma$\\
        &   [GeV]                          & $[10^{-12} \rm ergs \, cm^{-2} \, s^{-1}]$ & & \\
  \hline
  \noalign{\smallskip}
  1 & 0.1--0.4 & $4.4\pm1.5 $ & 36 & 1.9$\pm$0.6 \\
  2 & 0.4--1.6 &  $3.9\pm0.5$ & 126 &  1.9 $\pm$ 0.3 \\
  3 & 1.6--3.0 &  $2.1\pm0.3$ & 105 & 2.7$\pm$0.5 \\
  4 & 3.0--10 &  $2.8 \pm 0.3$ & 87 & 1.47$\pm$ 0.05 \\
  5 & 10--100 & $1.6\pm0.4$  & 12 & 2.29 $\pm$0.06 \\
\noalign{\smallskip}
\hline 
 \end{tabular}
\label{fermi}
}

\end{table}

\section{Spectral energy distribution modelling}

In order to derive a spectral energy distribution (SED) over the whole electromagnetic waveband, we included radio to UV data from the literature. We used the {\it INTEGRAL} IBIS/ISGRI upper limit between 20--60 keV, the {\it INTEGRAL}/JEM-X detection between 3--10 keV and the 10--25 keV upper limit together with the {\it Fermi}/LAT data to represent the long-term average emission,
where we assume that the majority of the low-flux hard X-ray emission is due to the core rather than to the jet knots.
We combined these data with core detections from radio to infrared from NED \footnote{http://ned.ipac.caltech.edu/} and VHE data from H.E.S.S., which has observed M~87 between 2003 and 2006 \citep{Aharonian2006}.To model the 2006 {\it Suzaku} detection we use simultaneous observations, as described in Section 3.2.

To represent the broad-band SED we applied a one-zone synchrotron self-Compton (SSC) model. This model assumes an isotropic population of high-energy electrons that emit synchrotron radiation followed by inverse Compton scattering of the synchrotron photons to higher energies \citep{Maraschi1992}. 
In this simplified model the electron population is contained in a spherical volume with radius $R$ and a randomly orientated magnetic field $B$. The volume moves relativistically with a bulk Lorentz factor $\Gamma_b$ towards the observer in a jet with an angle $\theta$ to the line of sight. The emission is Doppler-shifted with a Doppler factor $\delta=[\Gamma_b (1-\beta \cos\theta)]^{-1}$. The electron energy distribution in the jet-frame is assumed to follow a broken power-law, with index $p_1$ between the minimum energy $E_{\rm min}$ and the break energy $E_{\rm br}$ and index $p_2$ between $E_{\rm br}$ and the maximum energy $E_{\rm max}$:
\begin{equation}
N(E)= 
\begin{cases}
k E^{-p_1}\, \text{if}& E_{\rm min} <E< E_{\rm br}  \\
k E^{-p_2}\, \text{if}& E_{\rm br}  <E< E_{\rm max}
\end{cases} 
\end{equation}
Here, $k$ is the electron normalisation factor, and $p_1<3$ and $p_2>3$. The peak frequencies are dependent on the break energy 
via 
\begin{equation}
\nu_s=\frac{4}{3}\gamma_{br}^2\frac{eB}{2\pi m_e c}\frac{\delta}{1+z} 
\end{equation}
for the synchrotron peak, where
\begin{equation}
\gamma_{br}=\frac{E_{br}}{m_e c^2}
\end{equation}
and for the frequency of the inverse Compton peak (in the Thomson regime)
\begin{equation}
\nu_{IC}=\frac{4}{3} \gamma_{br}^2 \nu_s
\end{equation}
where we assume that the dominant synchrotron power is emitted at the peak of the synchrotron branch. Here, $\gamma$ is the Lorentz factor of the relativistic electrons in the plasma blob.

For some objects, such as bright flat spectrum radio quasars (FSRQ), the SSC model does not properly represent the SED. 
In addition to the inverse Compton scattering of synchrotron photons, external seed photons from e.g. the broad-line region, that are Compton upscattered to higher energies can contribute to the inverse Compton emission \citep[external Compton component,][]{Dermer1993}. Otherwise, multiple scatterings should be considered to explain the Compton dominance of FSRQs \citep[e.g.][]{Georganopoulos2006}.

For the modelling of M~87 we have used a single-zone SSC code developed by A. Tramacere\footnote{http://www.isdc.unige.ch/sedtool/} (to be released soon) where the least-square method is used to find the best fit of the numerical modelling \citep{Massaro2006,Tramacere2009,Tramacere2011}. 
This code allows to apply several different electron energy distribution shapes, and we choose the broken power law shape to compare the result of the fitting with previous works. In addition to the SSC emission there is also a host galaxy component in the model, which shows a peak in the optical that is consistent with the data. The flux contribution of the host galaxy is $\nu f_{host} = (4\pm2)\times 10^{-12} \rm \, erg \, cm^{-2} \, s^{-1}$ for all fits presented below. 
In the following sections the results of the modelling is described (see Table~\ref{ssc_param} for the best-fit results). 

\begin{table*}
\centering
\caption{The results of the SED fitting of M~87. The first three columns show the results of fitting the average low-flux state using several different configurations for the jet angle. The columns "2006 core" and "2006 HST-1 knot" refer to the 2006 SEDs of the M~87 core and HST-1 jet knot, respectively. The last column shows the result of the SED model presented by \citet{Abdo2009}. }
{\footnotesize
\begin{tabular*}{\textwidth}{l|l|l|l|l|l|l}  
\hline \hline
\noalign{\smallskip}
Parameter          & $\theta=15^{\circ}$& Fixed beam          & $\theta=10^{\circ}$ & 2006 core& 2006 HST-1 knot & Abdo 2009$^{c}$\\
  \hline
  \noalign{\smallskip}
$\theta$           &  $15^{\circ \, a}$ & -$^{b}$             & 10$^{\circ \, a}$   & $15^{\circ \, a}$ & $15^{\circ \, a}$  & $10^{\circ \, a}$ \\
$B [G]$            & $2.0\times10^{-3}$ & $1.7\times 10^{-2}$ & $2.2\times 10^{-2}$ & $3.3\times10^{-3}$& $6.4\times10^{-1}$ & $5.5\times 10^{-2}$   \\
$R [cm]$           &$5.6\times10^{17} $ & $3.3\times 10^{16}$ & $2.4\times10^{16}$  & $5.0\times10^{17}$& $1.0\times10^{16}$ &  $1.4 \times 10^{16}$  \\
$\Gamma_b$         &     $3.8 $         & -$^{b}$             & $3.4 $              &  $3.9$            & $1.2$              & $2.3$\\
$E_{\rm min} [eV]$ & $2.8\times10^7$    & $2.6\times10^{8}$   & $2.1\times10^8$     &  $6.7\times10^7$  & $6.0\times10^7$ &  $5\times 10^{5}$   \\ 
$E_{\rm max} [eV]$ & $2.3\times10^{13}$ & $6.5\times10^{13}$  & $5.1\times10^{13}$  & $2.5\times10^{13}$ $^{a}$& $2.5\times10^{13}$ $^{a}$ &   $5 \times 10^{12}$\\
$E_{\rm br} [eV]$  & $2.0\times10^8 $   & $1.3\times 10^9$    & $1.3\times 10^9$    & $5.0\times10^8$   & $1.0\times10^{8}$ &  $2\times 10^{9}$\\
$p_1$              &   $-1.8$            & $1.1$               & $1.1$               &    $2.8$          & $2.5$   &$1.6$ \\
$p_2$              &    $3.4$           & $3.5$               & $3.5$               &    $3.6$          &  $3.4$  &$3.6$\\
$\chi_\nu^2$       & 1.9\, (17 d.o.f.)  & 2.8\, (17 d.o.f.)   & 3.2\, (16 d.o.f.)   & 3.1\, (2 d.o.f.)  &  1.9\, (2 d.o.f.) &-\\

\noalign{\smallskip}
\hline \hline
 \end{tabular*}
 {\it a)} parameter fixed;\\
{\it b)} beaming factor $\theta$ fixed to 5, see text;\\
{\it c)} see \citet{Abdo2009}
\label{ssc_param}

}
\end{table*}

\subsection{SSC model for the average low-flux state}

At first, we kept the angle $\theta$ and the bulk Lorentz factor $\Gamma_b$ free with a fixed beaming factor $\delta = 5$. This value is consistent with the apparent motion of $\sim 0.5c$ observed in the jet \citep{Kellermann2007}. The fit has a $\chi^2$ of 2.8 (17 d.o.f., Table~\ref{ssc_param}). 
This gives a value of $E_{\rm min}=2.6\times 10^{8}\rm\, eV$ for the minimum energy of the electrons, $E_{\rm max}=6.5\times10^{13} \rm \,eV$ for the maximum energy and $E_{\rm br}=1.3\times 10^9\, \rm eV$ for the break energy. The magnetic field has a value of $B=1.7 \times 10^{-2}\rm\,G$ and the radius of the emission region $R=3.3\times 10^{16}\rm\,cm$. The indices of the broken power law are $p_1=1.1$ and $p_2=3.5$.  

Next we have fixed the jet angle at $\theta=10^{\circ}$, a value that is closer to that of blazars, although the apparent angle as seen in the large scale radio jet appears to be rather $\theta=15^{\circ}$. This allows us to compare our results of the SED fitting with the previous work of \citet{Abdo2009}, who also used $\theta=10^{\circ}$ for their SED fitting, although this value does not use the true orientation of the jet \citep[e.g.][]{Meyer2013}.
The fit gives a reduced $\chi_\nu^2 = 3.2$  (16 d.o.f., Table~\ref{ssc_param}). The magnetic field is $B= 2.2\times 10^{-2} \rm \, G$ and the radius of the emission region $R=2.4\times10^{16} \, \rm cm$. A bulk  
Lorentz
factor $\Gamma_b =3.4 $ is found, which is slightly higher than the bulk factor derived by \citet{Abdo2009} of $\Gamma_b=2.3$. 
The minimum energy is $E_{\rm min}=2.1\times 10^8\rm\,eV$, the maximum energy is $E_{\rm max}=5.1\times 10^{13} \rm \, eV$, the break energy $E_{\rm br}=1.3\times 10^9 \rm \, eV$, and the power law indices found are $p_1=1.1$ and $p_2= 3.5$, all consistent with the previous result using a free angle. Except for the minimum energy of the electrons these values are also consistent with the values used by \citet{Abdo2009}. 

\begin{figure} 
 \includegraphics[width=9cm, keepaspectratio=true]{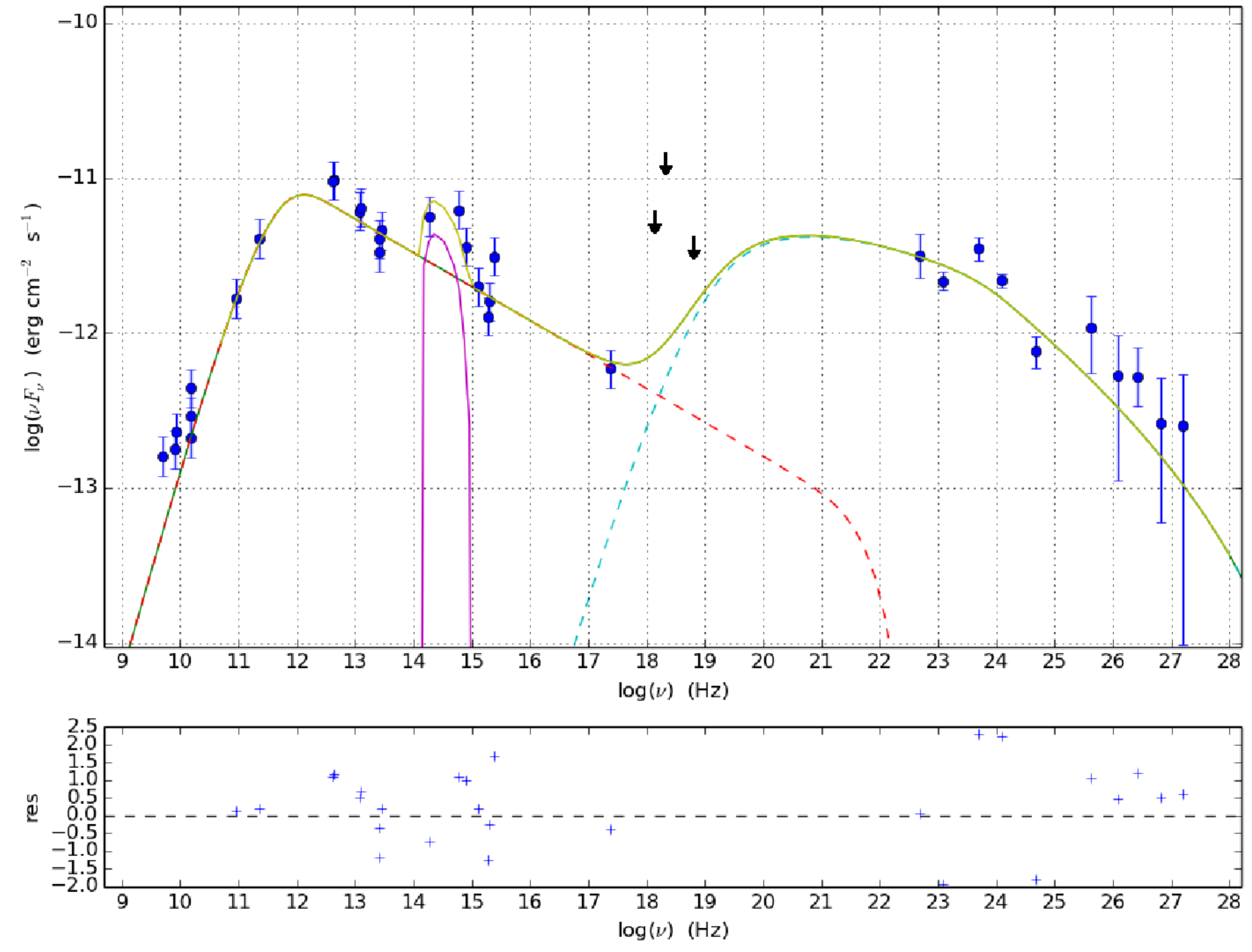}
 \caption{SED of the average low-flux state of M~87, using a jet angle of $\theta=15^{\circ}$.  The arrows show the {\it INTEGRAL} upper limits. The solid line from the radio to VHE domain shows the SSC fit. The synchrotron contribution, dashed, is dominant at low frequencies and the inverse Compton contribution, also dashed, is dominant at high frequencies. The host galaxy component is shown with a solid line in the range $10^{14}-10^{15}\rm\, Hz$. The {\it Chandra}/ACIS observation of the M~87 core at $2.4\times 10^{17}\rm\, Hz$ was taken at the end of 2008, when M~87 was in a quiescent state \citep{Abdo2009}.}
 \label{averageSED}
\end{figure}

After this we fitted the data with a more conservative angle of $\theta=15^{\circ}$, consistent with proper motion observations by \citet{Biretta1999} and \citet{Meyer2013}. The SED is presented in Fig.~\ref{averageSED}. The fit, with a reduced $\chi_\nu^2 = 1.9$  (17 d.o.f., Table~\ref{ssc_param}), yields a lower magnetic field, \mbox{$B=2.0\times10^{-3}  \rm \, G$}, and a larger 
emitting region, \mbox{$R=5.6\times 10^{17}\rm\,cm$}, compared to the previous fits.
 While the second index of the broken power law is similar with $p_2=3.6$, the first index changes to $p_1=-1.8$. The bulk Lorentz factor increased slightly to $\Gamma_b =3.8$. The minimum energy of the electrons decreased to $E_{\rm min}=2.8 \times 10^7\rm\,eV$ and the maximum energy 
 decreased slightly to $E_{\rm max}=2.3 \times 10^{13} \rm \,eV$. Also the break energy of the electron energy distribution decreased to $E_{\rm br}=2.0 \times10^8\, \rm eV$.

\subsection{SSC model for the high flux state in 2006}
Since the {\it Suzaku}/PIN data taken in 2006 indicate a flux level that is 3 times as high as the upper limit determined from the average 2003-2011 data provided by {\it INTEGRAL} IBIS/ISGRI, and similarly the {\it Suzaku}/XIS data show a 3--10 keV flux about four times that of the {\it INTEGRAL} JEM-X core flux, we also collected additional 2006 data in order to derive a simultaneous SED for this period.

Because we are not able to resolve the core and jet with  {\it Suzaku}/PIN, it is not clear whether the flux increase originates in the core or in one of the jet knots (see e.g. the light curves presented by \citealt{Abramowski2012}). Therefore we create two different SEDs: one for the M~87 core, and one for the bright jet knot HST-1, which is known to flare. We add simultaneous observations for both components from {\it HST} \citep{Perlman2011}, {\it VLA} \citep{Harris2009} and {\it VLBA} \citep{Cheung2007}. In addition we add simultaneous VHE data from {\it MAGIC} \citep{Berger2011}, however also in gamma-rays it is not possible to distinguish between the core and the HST-1 knot. 
Lastly, we also include the {\it HESS} observations from the average SED \citep{Aharonian2006}.
While these observations were not taken simultaneously, in 2006 no increased emission was observed from M~87.

\begin{figure} 
 \includegraphics[width=9cm, keepaspectratio=true]{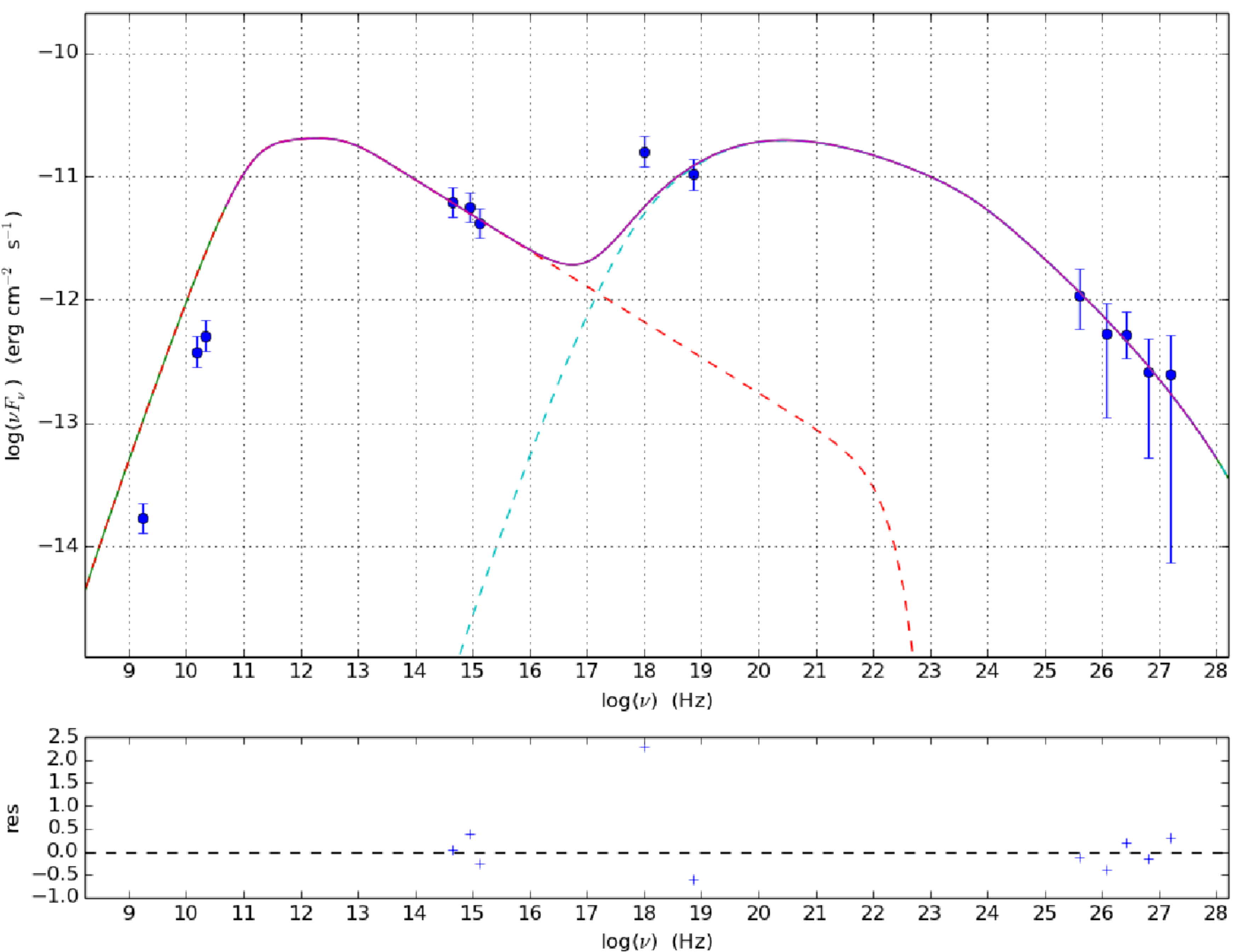}
 \caption{SED of the core of M~87 during the high-flux state in 2006, using a jet angle of $\theta=15^{\circ}$. The SSC fit is shown with a solid line. The synchrotron branch is shown with a dashed line at low frequencies, and the inverse Compton contribution is shown with a dashed line at high frequencies.} The Suzaku/PIN data do not seem to match the SED.  
 \label{2006core} 
\end{figure}

The SED for the M~87 core is presented in Fig.~\ref{2006core}. The fit has a reduced $\chi_\nu^2 = 3.1$  (2 d.o.f., Table~\ref{ssc_param}), where the maximum energy has been fixed to $E_{\rm max}=2.5\times10^{13} \rm \,eV$. Using an angle of $\theta=15^{\circ}$, the bulk Lorentz factor is consistent compared to the low-flux average state with $\Gamma_b =3.6$. The radius of the emitting region of $R=5.0\times 10^{17}\rm\,cm$ is consistent with the low-flux state modelled with the same jet angle, whereas the magnetic field slightly increased to $B=3.3\times10^{-3}  \rm \, G$. Both the minimum energy and the break energy of the electron energy distribution increased slightly to $E_{\rm min}=6.7\times10^7\rm\,eV$, and $E_{\rm br}=5.0\times10^8\, \rm eV$. The indices of the power law describing the electron energy distribution have indices $p_1=2.8$ and $p_2=3.6$. As can be seen in  Fig.~\ref{2006core}, the model is not consistent with the {\it Suzaku} spectrum, which dictates a steep power law in the X-ray regime. Therefore, the {\it Suzaku} spectrum implies that the hard X-ray data describe rather the tail of the synchrotron branch than the inverse Compton branch.

\begin{figure} 
 \includegraphics[width=9cm, keepaspectratio=true]{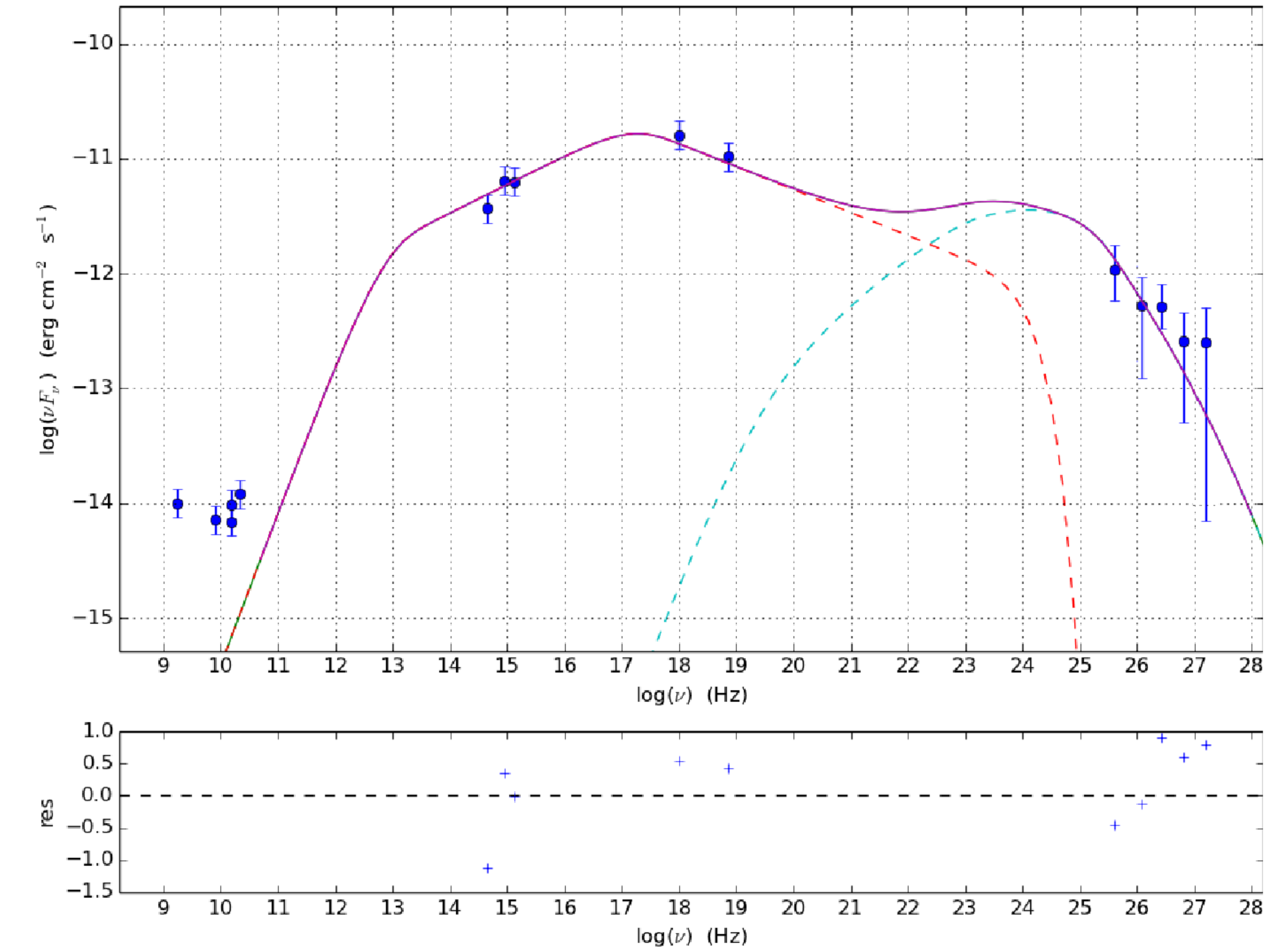}
 \caption{SED of the HST-1 knot of M~87 during the high-flux state in 2006, using a jet angle of $\theta=15^{\circ}$.  The SSC fit is shown with the solid line. The synchrotron branch, dominant at low frequencies, and the inverse Compton contribution, which is dominant at high frequencies are shown with dashed lines.} The Suzaku/PIN data match the SED.
 \label{2006knot}
\end{figure}
Since the jet-knot HST-1 is known to flare we also model it as a potential origin of the hard X-ray emission detected by {\it Suzaku}, see Fig.~\ref{2006knot}. A jet angle of $\theta=15^{\circ}$ is used and the fit has a reduced $\chi_\nu^2 = 1.9$  (2 d.o.f., Table~\ref{ssc_param}). Also in this case the maximum energy is fixed to \mbox{$E_{\rm max}=2.5\times10^{13} \rm \,eV$} to allow convergence. The bulk Lorentz factor 
is lower compared to the 2006 core fit and the average low-flux state with $\Gamma_b =1.2$. The 
size of the emitting region is also 
smaller, 
with $R=1.0 \times 10^{16} \rm \, cm$. The magnetic field strength increased to $B=6.4\times10^{-1}  \rm \, G$. The minimum energy of the electrons is similar to the fit of the core with $E_{\rm min}= 6.0 \times 10^7 \rm \, eV$, and the break energy decreased to $E_{\rm br}=1.0\times10^8\rm\,eV$. The power law indices that characterise the electron energy distribution are similar to the 2006 core fit with $p_1=2.5$ and $p_2=3.4$. As can be seen in Fig.~\ref{2006knot}, the {\it Suzaku} data points lie on the inverse Compton branch, rather than on the synchrotron branch, consistent with the hard X-ray spectrum.

\section{Discussion}

In the following we first discuss the possible origin site of the emission seen in {\it INTEGRAL}/JEM-X, before we turn to the hard X-ray variability and the spectral energy distribution. Finally, we discuss the possible source type at the origin of the high-energy emission in M~87.

\subsection{JEM-X: core emission}
Due to the the angular resolution of JEM-X (FWHM = 3 arcmin), the flux of M~87 observed by JEM-X is the sum of core, jet and extended diffuse emission. Here we are interested only in the core emission, because we want to investigate the origins of the high-energy emission, which is emitted from, or close to the core. In addition, the measured JEM-X flux is an average over several years. 
The jet knot HST-1 is known to be variable \citep{Harris2009}, and was in outburst between 2003 to 2007, with the luminosity of the knot peaking in 2005.  For a light curve of the X-ray emission of HST-1 see Fig.~1 in \cite{Abramowski2012}. Since the JEM-X observations we used are taken mostly after 2008, and none during 2005, we assume that we did not include data during which the knot was in outburst.  

Using literature values from observations of the knot HST-1 made with {\it Chandra}/ACIS by \citet{Perlman2005} and by \citet{Wilson2002} for other jet knots, we find that the combined jet flux is about \mbox{$f(3-10 \rm \, keV) = 10^{-12} \rm \, erg \, cm^{-2} \, s^{-1}$}. In quiescence the jet emission is dominated by the jet knots HST-1, A, and D.
For the extended emission we used observations from {\it Chandra}/ACIS, taken beginning of May 2005. We analysed the data, with a total exposure of 123 ks, and extracted a spectrum in a circular region with a radius of 2.5 arcminutes centred around the core of M~87. The core and jet were excluded in the analysis. This spectrum yielded a flux of $f(3-10 \rm \, keV) = 9\times 10^{-12} \rm \, erg \, cm^{-2} \, s^{-1}$. This implies that the core emission is about $f= 6 \times 10^{-12} \rm \, erg \, cm^{-2} \, s^{-1}$, with a variability by a factor of two in X-rays \citep{Hilburn2012}. 

The spectrum extracted from the 2006 {\it Suzaku}/XIS observation also showed a thermal and non-thermal component due to the extraction radius covering both the core and surrounding medium. Since a spectrum is available, disentangling these components is a more straightforward task and we find a 3--10 keV flux of \mbox{$f= 2 \times 10 ^{-11}\rm \, erg \, cm^{-2} \, s^{-1}$} for the non-thermal component, which is almost four times the JEM-X core flux. The strongly increased flux derived from the XIS data implies a flare when compared to the average flux.

\subsection{Hard X-ray variability}

 We found an overall 3$\sigma$ upper limit to the M~87 flux of $f(20 - 60 \rm \, keV) < 3 \times 10^{-12} \rm \, erg \, cm^{-2} \, s^{-1}$ using all available {\it INTEGRAL} data for a total effective exposure time of 1.7~Ms. 
Apart from a likely spurious detection by \mbox{{\it INTEGRAL}} at 5.1$\sigma$, corresponding to a flux of \mbox{$f(20-60 \rm \, keV) = (9 \pm 2) \times 10^{-12} \rm \, erg \, cm^{-2} \, s^{-1}$} \citep{Walter2008}, M~87 had not previously been detected in hard X-rays. Using the same data set as \citet{Walter2008} but applying the latest analysis software OSA~10 we derived a $3\sigma$ upper limit flux of $f(20-60 \rm \, keV) < 8 \times 10^{-12} \rm \, erg \, cm^{-2} \, s^{-1}$, inconsistent with the earlier detection claim.  

This does not mean, however, that the hard X-ray flux of M~87 never exceeds this average value. We found that, during a {\it Suzaku}/PIN observation at the end of November 2006, M~87 is detected with a flux of \mbox{$f(20-60 \rm \, keV) = 10^{-11}\rm \, erg \, cm^{-2} \, s^{-1} $}. There is no \mbox{{\it INTEGRAL}} IBIS/ISGRI observation at this time to confirm the detection, but in July 2006 {\it INTEGRAL} has observed M~87 for 23 ks constraining a $3\sigma$ upper limit of $f (20 - 60 \rm \, keV) < 3\times 10^{-11}$  $\rm \, erg \, cm^{-2} \, s^{-1}$, consistent with the flux measured by {\it Suzaku}/PIN.

\subsection{Spectral energy distribution}

To characterise the broad-band emission of M~87 we have produced three SEDs; one representing the average low-flux state and two representing the increased X-ray emission as observed by {\it Suzaku} at the end of 2006. The data used in the average low-flux SED have been taken at different times, which might result in a biased measurement in case of variability since several spectral states might contribute to the data.

To investigate whether there is contamination of different spectral states in the M~87 SED, we consult light curves of M~87 from 2001 to 2011 based on radio, optical, X-rays and VHE observations \citep{Abramowski2012}. The VHE band shows increased activity in 2005, 2008 and 2010, during which the X-ray emission also increased by a factor of 3. During the VHE flare in 2008 the radio emission from the jet base increased as well, indicating the VHE emission is likely produced near the SMBH. However, during the VHE flare in 2008, no enhanced radio emission has been observed. The VHE observation used in the average low-flux SED was taken in 2004 when M~87 was in a low state, so there is no contamination from any of the observed VHE flares. 
The radio and optical detections of the core used in the SED are averaged over several years, and the light curves show variations by a factor of $\sim 2$. Variability at this level will not significantly alter the physical parameters derived from modelling the SED. A similar approach of using time-averaged SEDs has been applied already successfully to M~87 \citep{Abdo2009} and other radio galaxies, for example Pictor~A \citep{Brown2012} and 3C~111 \citep{deJong2012}. 

The high flux detected by {\it Suzaku}/PIN indicates a flaring state of the source. The phenomenology of the flaring activity of M~87 hints at a behaviour similar to that observed in blazars, and in particular in 
BL~Lac objects.
During a flare an increase of the peak energies of the SED correlates with an increase in the corresponding peak fluxes \citep[see for example Mrk~501,][]{Tavecchio2001}. In 2005, the increase in VHE, X-ray and optical emission coincided with the expulsion of the jet knot HST-1 from the core. During the {\it Suzaku} observations at the end of 2006 the light curves presented in \citet{Abramowski2012} do not show increased activity in the VHE or optical band.  Even though the amount of simultaneous observations is sparse, based on the 2006 SED modelling the HST-1 knot is the more likely candidate for the hard X-ray emission. 
The SED models show that the hard X-ray spectrum can not be reconciled with the simultaneous core observations. If the hard X-ray band would be on the tail of the synchrotron branch for the 2006 core observations, this would indicate a synchrotron peak frequency shift from $\sim 10^{12}\rm\,Hz$ for the low-flux average state to $>10^{16}\rm\, Hz$ for the flaring state. A peak frequency shift this large has not been observed even in the most extreme blazar flares. 
{\it Chandra} had also observed M~87 on 29 November 2005, but due to pile-up no useful spectra could be extracted during this time \citep{Harris2009}. Since {\it Chandra}/ACIS has a small angular resolution of 0.5'', it is possible to monitor the core and jet knot separately. Because of the pile-up during the observation, the intensity is expressed in detector-based units of $\, \rm keV\,s^{-1}$ where the entire energy band from 0.2-17 keV is integrated. For the core an intensity of $0.51\pm0.02 \, \rm keV\,s^{-1}$ was measured, whereas the HST-1 knot has an intensity of $4.03\pm0.04 \, \rm keV\,s^{-1}$. This supports our conclusion that the X-ray flare detected by {\it Suzaku} originates in the HST-1 knot rather than the core.  

Since similar SED modelling has been applied to other $\gamma$-ray bright radio galaxies, comparing the physical parameters derived from these models with the average fit of M~87 will help us put the results in a broader frame-work. Due to the degeneracy of several SED parameters, we can qualitatively compare the bolometric luminosity with the magnetic field $B$, the radius of the emitting region $R$ and the Doppler factor $\delta$, as these parameters influence the overall SED power.
 
The FR-I radio galaxy Cen~A has also been observed in the gamma-ray band, 
and similar to M~87 the overall SED can also be modelled by a simple SSC process \citep{Abdo2010CenAcore}, although a strong Seyfert contribution is visible in the X-ray and optical domain \citep{Beckmann2013}. \citet{Abdo2010CenAcore} present several SED fits to the core of Cen~A. Comparing the results for M~87 (jet angle $\theta=15\,^{\circ}$) to the model for Cen~A (jet angle  $\theta=30\,^{\circ}$), M~87 displays a slightly higher Doppler factor of $\delta=3.9$, compared to the Doppler factor used for Cen~A ($\delta = 1$) due to the smaller jet angle $\theta$. A lower Doppler factor causes the emission to appear less boosted. The magnetic field $B$ and radius $R$ of the emitting region are quite different for both sources: for Cen~A a value of $B = 6.2 \rm \, G$ was found for the magnetic field, and a much lower value of $B=2.0\times 10^{-3}\rm \, G$ for M~87. Since the synchrotron emission depends on $B$, a stronger magnetic field will result in a higher synchrotron flux. The radius of the emitting region is reported to be $R=5.6 \times 10^{17}\rm \, cm$ for M~87 
and $R=3 \times 10^{15} \rm \, cm$ for Cen~A \citep{Abdo2010CenAcore}. The radius $R$ defines the amount of radiating particles, therefore a larger emitting region results in a higher flux. The lower magnetic field strength $B$ used to model M~87, combined with the larger radius $R$ compared to Cen~A then results in a similar overall power in the SED, consistent with the comparable bolometric luminosity of about $L_{\rm bol}\sim10^{42}\rm\, erg \,s^{-1}$ of the two sources.

Comparing the SED parameters with the luminous ($L_{\rm bol}=5\times10^{44}\rm\, erg \,s^{-1}$) FR-II galaxy 3C~111 shows the differences between these two types of sources. A larger Doppler factor $\delta=14$, a smaller emitting region \mbox{($R=2\times10^{16}\rm\,cm$)} and larger magnetic field ($B= 0.04\rm\,G$) are used to model 3C~111 \citep{deJong2012}. The smaller volume of the emitting region used to model 3C~111 compared to M~87 yields a lower flux, but the strong Doppler factor $\delta$ and larger magnetic field of 3C~111 increase the flux strongly, consistent with the larger bolometric luminosity of this source compared to M~87.

As FR-I radio galaxies like M~87 are the parent population of BL~Lacs, comparing the resulting SED of M~87 with a BL~Lac object Mrk~421 will illustrate the differences between these two source classes. Using a one-zone SSC model, \citet{Blazejowski2005} found that Mrk~421 ($L_{\rm bol}=3\times10^{45}\rm\, erg \,s^{-1}$) requires a beaming factor of $\delta=10$, an emitting region a radius of $R=7.0\times10^{15}\rm\,cm$ and magnetic field $B = 0.405\rm\, G$. The stronger beaming, due to the small jet angle of Mrk~421, boosts the intrinsic emission strongly. While the emitting region is smaller compared to that used to model M~87, the magnetic field is much stronger, in combination increasing the flux.   

\subsection{M~87, a radio galaxy with a low luminosity BL Lac core}

In the overall context of SED models of $\gamma$-ray bright sources, we can compare the derived values with the average ones for Fermi/LAT detected BL~Lacs and FSRQs. 
As the FR-I radio galaxies can be understood as the parent population of the BL Lacs (e.g. \citealt{Urry1995}), also their SED parameters are closer to the ones found for BL~Lacs than for the brighter FSRQ class. As pointed out by \cite{Ghisellini2010}, the BL~Lacs on average appear to have similar masses as the FSRQs, with values around $10^8 - 10^9 \, \rm M_\odot$. The SMBH in M~87 exceeds this average value, and the bolometric luminosity $L_{\rm bol} \simeq 10^{41}- 10^{42} \rm \, erg \, s^{-1}$ results in an Eddington ratio of only $\lambda = L_{\rm bol}/L_{\rm Edd} \simeq 10^{-3}- 10^{-2}$ \citep{Owen2000,DiMatteo2003}. Nevertheless, a jet model similar to BL~Lacs can be applied to the M~87 SED.

Only 15 radio galaxies and a few misaligned blazars 
have been detected with {\it Fermi}/LAT, compared to more than 1500 observed blazars \citep{thirdagncat}. In the case of blazars the $\gamma$-ray emission is postulated to originate in the relativistic jet, which is pointed towards the observer. For M~87 the observed jet angle is $\theta=15\,^{\circ}$, causing the emission to be de-boosted as shown by the low Doppler factor ($\delta \sim 3.9$ ) of the emission. Even though the origin of the $\gamma$-ray emission is not completely clear, the modelling with a SSC model has shown that the high-energy emission is likely to arise in the jet.

BL Lac sources are thought to have an advection dominated accretion flow (ADAF; \citealt{Reynolds1996,DiMatteo2003})  rather than an accretion disk, which is present in FSRQs. Due to the ADAF, the accretion is less efficient, and as such causes the source to be less luminous compared to FSRQs. Since in the unified scheme FR-I radio galaxies, such as M~87 are considered to be misaligned BL Lacs, FR-I galaxies are thought to be powered by an ADAF as well. In the case of M~87 the broad-band emission is shown to be jet-dominated. 

The properties of M~87's central engine are closer to that of BL~Lacs than to FSRQs. The strength of the magnetic field, the bulk Lorentz factor of the jet and also the electrons' Lorentz factor distribution indicate rather a BL~Lac type emission. Also the overall bolometric luminosity of \mbox{$L_{\rm bol} \simeq 10^{41} - 10^{42} \rm \, erg \, s^{-1}$} \citep{Owen2000, DiMatteo2003} giving an Eddington ratio of only $\lambda \simeq 10^{-3} - 10^{-2}$ points in this direction. The observed physical properties of this AGN put it in a low luminous BL Lac class. The low power jet can be explained by the low accretion rate, rather than by the large mass, although the finding by \citet{Laor2000} that AGN with black hole masses larger than $M_{\rm BH} > 10^9 \rm \, M_\odot$  are radio-loud, seems to hold also in the case of M~87.

3C~120, another FR-I radio galaxy, was not included in the first, second or third {\it Fermi} catalogue, but using a 15-month data set \citet{Abdo2010misaligned} derived a significance of $\sim 5.6\sigma$ between 0.1--100 GeV for this source. {This source undergoes series of flares with a low long-term average flux \citep[e.g.][]{Sahakyan2014}. 3C~120 has been monitored in the radio, UV and X-ray band \citep{Lohfink2013}. From the X-ray observations by {\it Suzaku} and {\it XMM-Newton} a 0.4--10 keV spectrum is derived, where a fit with a composite jet+accretion disk model is favoured. The broad-band SED of this source has shown that the jet dominates the radio and $\gamma$-ray emission, and contributes only $\sim$10\% in the optical, UV and X-ray bands \citep{Kataoka2011}. The $\gamma$-ray spectrum is soft, with $\Gamma_{\rm \gamma}\sim2.7$, and the source is not detected in the TeV band, implying that the inverse Compton peak of this source is located in the X-ray to MeV band. In the case of 3C~120, as the $\gamma$-ray emission is variable on the time scale of months. \citet{Sahakyan2014} argue that the emission comes from a compact, relativistically moving emission region, but exclude the jet knots as the origin as these are deemed too large.

Only a few of the $\gamma$-ray detected radio galaxies have been observed in VHE. Recently the MAGIC telescopes have detected another radio galaxy, the FR-I IC~310 at energies $E>300\rm\,GeV$ \citep{IC310}. The SED of this source shows an inverse Compton peak in the TeV range. Combined with the low luminosity of this source, it was argued that IC~310 is an extreme case within the blazar sequence, with an extremely low accretion rate. Another possibility is that IC~310 is rather a misaligned version of an extreme BL~Lac, where the low luminosity is connected to the large viewing angle. This does, however not explain the observed TeV variability. M~87 is also observed in the VHE band, and shows an inverse Compton peak in the hard X-ray/soft $\gamma$-ray band. With its low bolometric luminosity, M87 is closer to IC~310 than to 3C~120, since both IC~310 and M87 have been detected firmly in both the $\gamma$-ray and TeV band. 

The low Eddington ratio  can be understood considering that the M~87 core is likely not fed by an accretion disk but by a radiatively inefficient accretion flow (RIAF) or advection dominated accretion flow (ADAF;  \citealt{Reynolds1996,DiMatteo2003}). The assumption that the accretion is radiatively inefficient also explains why in the case of M~87 we do not see a significant thermal inverse Compton component in the X-rays, and an optical spectrum consistent with that of a LINER. This is different to other $\gamma$-ray bright radio galaxies, which show an optical Seyfert core as e.g. Cen~A \citep{Beckmann2011} and 3C~111 \citep{deJong2012}.

\section{Conclusion}

We report, for the first time, a hard X-ray detection of the FR-I radio galaxy M~87 using {\it Suzaku}/PIN data. The observations were made between November 29 and December 2, 2006 with an elapsed time of 187 ks, resulting in a flux of $f = 10^{-11}\rm \, erg \, cm^{-2} \, s^{-1} $ between 20 and 60 keV. In addition, we derive a 3$\sigma$ upper limit of \mbox{$f(20-60 \rm \, keV) < 3\times 10^{-12} \rm \, erg \, cm^{-2} \, s^{-1}$} for the multi-year time averaged emission, based on 1.7 Ms of {\it INTEGRAL} IBIS/ISGRI data.

By modelling the broad-band energy distribution with a one-zone SSC model we connect the average hard X-ray upper limit to the emission and $\gamma$-ray emission to the core emission. The SED parameters show that M~87 can be considered to be a weak BL~Lac object, consistent with the advection dominated accretion flow model for the core \citep{Ptak1998,Reynolds1996} and the overall low-luminous FR-I nature of this galaxy. 
The high X-ray flux detected with {\it Suzaku} at the end of 2006 seems to indicate the source was undergoing an outburst or flaring episode. The steep slope of the spectrum, with a power law index of \mbox{$\Gamma=2.8^{+0.5}_{-0.4}$} between 20--60 keV, indicates that the emission was likely the high-energy tail of the synchrotron branch. Using simultaneous observations we created an SED for both the core and the jet knot HST-1 which is known to flare in other wavebands such as the radio and optical. Due the non-imaging character of the {\it Suzaku}/PIN observations, we are not able to determine whether the enhanced emission results rather from the core of M~87 or in the jet based on the X-ray observations alone, but from the SED modelling we conclude that the jet knot is the more likely candidate for the hard X-ray emission detected in 2006.

In the unification model for AGN, radio galaxies are the counterparts of blazars, where the lower luminosity BL~Lacs are linked to FR-I galaxies and the more powerful flat spectrum radio quasars (FSRQ) to the FR-II sources. One way to account for the differences in luminosity and Compton dominance between the two source classes is to model the SED of BL Lacs with a simple SSC model and the FSRQ with a more complex model, for example using an external Compton component in addition to the SSC \citep{Ghisellini1998}. The simple SSC model for the overall SED appears to be valid for this class of {\it Fermi}/LAT detected radio galaxies. Also in the cases of the brighter FR-II objects, like 3C~111, no external Compton component seems to be necessary to represent the SED, which is not in line with the unification model. Thus, in all these cases the dominating emission region is either far from a strong field of external seed photons, like the broad line region (as in the case of 3C~111), or the broad line region itself is weak because of radiatively inefficient accretion, as might be the case in M~87. 
The hypothesis, that EC is not significant in $\gamma$-ray detected radio galaxies should be tested in the case of the {\it Fermi}/LAT detected steep spectrum radio quasar 3C~207.0, which hosts a Seyfert 1.2 core. At a redshift of $z =0.68$ this object has a luminosity of $L(2-10 \rm \, keV) = 2.3 \times 10^{45} \rm \, erg \, s^{-1}$, and the strong Seyfert core that displays an iron K$\alpha$ line with $EW \simeq 60 \rm \, eV$ should give rise to a significant photon field able to provide ample seed photons for inverse Compton processes in this case.

\section*{Acknowledgments}
The authors thank Juan Antonio Zurita Heras, Fabio Mattana and Volodymyr Savchenko for their support in the {\it INTEGRAL} analysis and Katja Pottschmidt for her advice on the {\it Suzaku}/XIS data analysis. We also thank the referee Eric Perlman for the fruitful discussion and the advice that helped us to improve the manuscript.
This research is based on data provided by {\it INTEGRAL}, an ESA project funded by ESA member states (especially the PI countries: Denmark, France, Germany, Italy, Spain, Switzerland), Czech Republic, Poland, and with the participation of Russia and the USA. This research has also used data obtained from the Suzaku satellite, a collaborative mission between the space agencies of Japan (JAXA) and the USA (NASA). This research has made use of NASA's Astrophysics Data System Bibliographic Services. 
We acknowledge the financial support from the UnivEarthS Labex program of Sorbonne Paris Cit\'e (ANR-10-LABX-0023 and ANR-11-IDEX-0005-02) within the project ``Impact of black holes on their environment''. S.S. acknowledges the Centre National d'\'{E}tudes Spatiales (CNES) for financial support.

\bibliographystyle{mn2e_fix} 
\bibliography{M87.bib} 

\end{document}